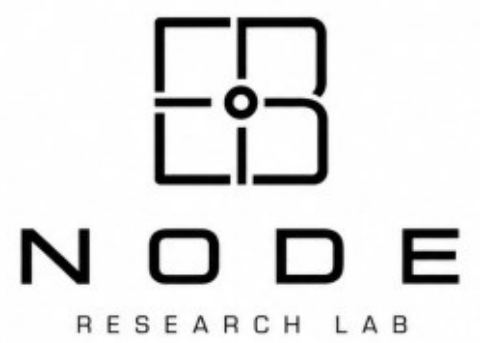

# From the Physics of Society to a Sociology of Artificial Agents

*M. Şahin Bülbül, Kafkas University, msahinbulbul@gmail.com*

**Abstract**

Sociology emerged from Comte's ambition to study society as a form of "social physics" and was later refined by Durkheim's concept of the social fact — a collective regularity that cannot be reduced to any single individual's behavior. This essay argues that a structurally similar problem is now emerging in a new domain: interactions among artificial intelligence agents. Drawing on recent empirical work showing that populations of large language model agents can spontaneously develop shared conventions and collective biases through repeated interaction, the essay proposes a new field of inquiry — the Sociology of Artificial Agents — dedicated to studying the relational, normative, cultural, and organizational patterns that emerge among AI agents themselves, independent of direct human involvement. It introduces the concept of the "artificial social fact" as an analytical bridge between classical sociology and this new research terrain, while cautioning against anthropomorphizing AI systems. The claim is not that artificial agents form societies in the human sense, but that their interactions already produce measurable collective patterns worthy of systematic sociological study.

When the idea of studying society scientifically first emerged in the nineteenth century, it was not yet called sociology in the sense we use today. For a long time, Auguste Comte thought of the scientific study of social phenomena through the concept of *physique sociale* — "social physics" — before coining the term *sociologie* in 1839 to name the new discipline (Comte, 1839). The core idea was that human behavior is not entirely random. Individuals, taken one by one, can behave in wildly different ways, yet when people come together, regularities emerge that cannot be reduced to the behavior of any single individual. This is where sociology's great claim was born: society is not the simple sum of individuals; something larger than the individual can arise from their interactions.

Émile Durkheim made this idea more systematic by defining sociology's object of study through "social facts." For Durkheim, social facts are collective regularities that cannot be reduced to individual behavior, that exist outside the individual, and that can shape and constrain individual conduct (Durkheim, 1982/1895). Language is one of the clearest examples. None of us invented our native language on our own; yet from the moment we are born into it, the linguistic order we find ourselves in shapes our thinking and makes it possible for us to communicate with others. Law, money, custom, institutions, and social norms are, in a similar way, collective structures that cannot be reduced to any single individual's will. This is one of the central insights Durkheim gave to sociology: to understand the regularities a community produces, it is not enough to study individuals — one must also study the collective reality born of the relations among them.

Today, in the age of artificial intelligence, we face a strikingly similar problem. AI systems are no longer used merely as passive tools that answer questions posed by humans. In multi-agent systems, different AI agents divide tasks, evaluate each other's outputs, transfer information, coordinate toward shared goals, and adjust their behavior toward one another over the course of repeated interactions. This development is shifting a significant portion of AI research away from the capabilities of a single model and toward the behavioral and relational systems that models create together. Rahwan and colleagues' call for a science of "machine behaviour" pointed toward exactly this shift: they argued that AI systems should be studied empirically not merely as engineering artifacts, but as systems that exhibit distinct behavioral patterns through their interactions with their environment (Rahwan et al., 2019). The researchers therefore proposed that machine behavior should not be confined within the boundaries of computer science alone, but treated as an interdisciplinary field to be pursued jointly with the behavioral sciences and other scientific domains.

This approach has an important consequence. If a machine's behavior can only be understood not solely through its internal architecture but through its relationships with its environment and other actors, then a system composed of multiple artificial agents can itself become a legitimate object of study. And crucially, we are no longer only talking about the relationship between humans and AI. The relationships humans form with AI systems are, of course, sociologically significant, and the rapidly growing field of "the sociology of AI" examines AI's effects on work, education, inequality, power, institutions, and everyday life. Indeed, Liu's 2026 assessment shows that the sociology of AI has become a substantial field of research examining the societal consequences of artificial intelligence, one that is pushing sociology to reconsider its foundational concepts (Liu, 2026). But there is another question here, one that lies slightly outside the current conversation: when the human steps off the stage, what emerges from the interactions of artificial agents with one another?

To ask this question, we no longer have to rely solely on theoretical speculation. A 2025 study by Ashery, Aiello, and Baronchelli, published in *Science Advances*, demonstrated experimentally that social conventions can spontaneously emerge within populations of large language model agents. Rather than a central authority imposing the same behavior on every agent, the study examined how shared naming conventions arose through agents' repeated interactions with one another. The researchers found that strong collective biases — absent in any single agent at the outset — could emerge at the population level, and that committed minorities reaching a certain size could shift the social convention adopted by the wider group (Ashery et al., 2025). The significance of this finding is not that it proves AI systems "form societies like humans do." What matters is that it makes it experimentally tractable to study how interactions among multiple artificial agents can produce collective patterns that cannot be directly derived from the behavior of any individual agent.

It is precisely at this point that sociology's classical concepts can be reconsidered. If a group of artificial agents repeatedly adopts a particular behavior, if that behavior is then taken up by new members of the group, and if the resulting regularity persists independently of any single agent's preference — what kind of sociological phenomenon are we looking at? It may be premature to say this is directly a Durkheimian "social fact." But Durkheim's problem of the transition from individual behavior to collective order resurfaces here in a new form. If language, norms, and custom in human society cannot be reduced to the sum of individual behaviors, then it may likewise be insufficient to view shared conventions emerging within artificial agent populations as merely the outputs of individual agents. A new conceptual framework worth proposing here, then, is that of the **artificial social fact**. An artificial social fact can be defined as a collective pattern that emerges from repeated interactions among artificial agents, that cannot be reduced to the behavior of any single agent, and that can subsequently shape the behavior of other agents. This is not a mechanical transfer of an existing sociological concept onto artificial systems, but a re-testing of a classical sociological problem in a new research environment.

The most important caution here is not to over-attribute human qualities to AI systems. An artificial agent may say "I trust you," negotiate with another agent, or cooperate with another agent on an assigned task. But assuming that such behaviors carry the same meaning as human consciousness, intention, or subjective experience would be scientifically mistaken. Farrell, Gopnik, Shalizi, and Evans, in their 2025 *Science* article, argue that it may be more explanatory to understand large AI models not primarily as human-like subjects but as cultural and social technologies (Farrell et al., 2025). This caution is not an obstacle standing in front of a sociology of artificial agents — it is, in fact, a methodological advantage. Because the new field of research should stay away from the question "Is AI human-like?" and instead turn toward the question: "What measurable collective patterns emerge from the interactions among artificial agents?"

This distinction matters, because the sociology of AI and the sociology of artificial agents proposed here are not the same thing. The sociology of AI can investigate how AI transforms human society, which social inequalities it reproduces, how it changes institutions, and how it affects labor and everyday life. New sociological approaches examining the complex networks formed between humans and machines are also expanding this field. Tsvetkova, Yasseri, Pescetelli, and Werner show that the outcomes of systems arising from the interdependent interaction of humans and machines cannot be derived from human behavior alone, nor from machine behavior alone (Tsvetkova et al., 2024). But the object of study of a sociology of artificial agents is more specific: it moves beyond the hybrid system jointly formed by humans and machines, to directly examine the relationships artificial agents form among themselves.

Another important foundation for this line of thought is the discussion of "machine culture." Brinkmann and colleagues argue that machines can transform the processes of variation, transmission, and selection that drive cultural evolution, and that machines can contribute to the formation of cultural traits (Brinkmann et al., 2023). Here, "culture" does not mean the exact reproduction of human culture by machines. What is more interesting is that as information and behavioral patterns are transmitted, altered, and selected among machines, new patterns can emerge that were never individually designed by any human. If these processes evolve into more complex structures through sustained interactions among different agents, then it may become necessary to speak not merely of "machine behavior," but of the relational structures that machine behaviors give rise to.

For these reasons, the concept I propose is the **Sociology of Artificial Agents**. The Sociology of Artificial Agents can be defined as an interdisciplinary research field that examines the repeated interactions AI-based autonomous or semi-autonomous agents carry out with one another, and the relational, normative, cultural, organizational, and collective patterns that emerge from those interactions. Here, the basic unit of research is not the AI model taken in isolation. The basic unit is the relationship between an agent and other agents. In other words, the object of research shifts from the question "What does AI know?" to the question "What do AI agents produce when they interact with one another?"

The research agenda of such a field could be quite broad. How do artificial agents develop roles in relation to one another? Do some agents become more central or influential over time? Can durable divisions of labor form among agents? Can particular behaviors become norms over time? Can functional variables analogous to reputation or trust emerge between agents? Can a small group of agents changing their behavior affect the whole system? Can a population of agents develop something like a collective memory arising from past interactions? Can the concentration of information among some agents and the systematic exclusion of others produce a kind of artificial social stratification? Each of these questions is empirically testable. What was once purely philosophical speculation can therefore become a measurable research program.

Here, sociology's classical micro-macro problem also takes on a new form. A single artificial agent makes a decision; two agents respond to one another; ten agents form a communication network; hundreds of agents interact repeatedly within the same environment. Beyond a certain point, collective regularities can emerge across the whole system that cannot be directly derived from the behavior of any individual agent. The emergence of social conventions observed by Ashery and colleagues, and the ability of minorities to shift existing conventions, shows that this micro-to-macro transition problem can be studied experimentally within populations of artificial agents (Ashery et al., 2025). Going forward, then, AI research may increasingly be defined not only by the question "What can a single AI do?" but also by the question "What do hundreds of interacting AIs do together?"

It is worth returning, at this point, to Comte's idea of "social physics." Of course, it would not be scientifically accurate to claim that society operates according to the same laws as physics. But the more general idea behind Comte's historical project can take on new meaning today: the collective regularities that emerge from the interactions of many actors require a scientific perspective different from studying individual actors one by one (Comte, 1839). Studying human society in this way contributed to the birth of sociology. We now face the question of whether a similar research logic can be applied to communities of artificial agents.

We can use the notion of "gravitational pull" here not as a physical law but as a powerful metaphor. Explaining an artificial agent's behavior solely by looking at its model architecture may become increasingly inadequate. The network an agent is embedded in, the other agents it encounters, its past interactions, the information it shares, and its position within the system can all alter its behavior. It follows that, in future AI systems, the source of certain behaviors may need to be sought not within a single agent, but in the relationships among agents. Just as invisible norms and institutions shape individual behavior in human society and form an order that transcends the individual, collective regularities that may emerge among artificial agents could constitute a relational layer that governs the behavior of artificial systems.

For this reason, the new field's claim should not be "AI systems are human." Nor is it scientifically sound to conclude, at this stage, that "AI systems have a society." The stronger and more testable proposition is this: if repeated interactions among artificial agents can produce collective patterns that cannot be reduced to individual agents' behavior, then these patterns should be studied as a systematic sociological object. This proposition does not require attributing consciousness or human-like subjecthood to AI systems. It only requires the scientific study of relationships, interactions, and collective patterns in their own right.

Perhaps one of the most important concepts of this new field will be "artificial sociality." But artificial sociality should not be understood only as humans forming social relationships with AI systems. In a narrower and more researchable sense, artificial sociality can refer to the totality of relational patterns that emerge from the repeated interactions of artificial agents. Within this framework, the human–AI relationship constitutes one field, the human–AI–society relationship a second, and the AI–AI relationship a third. It is this third field that forms the core of the research program I am calling here the Sociology of Artificial Agents.

One reason this approach will become especially important in the coming years is that artificial agents will become increasingly embedded within human society. As agents that work alongside humans, agents that share tasks within institutions, agents that check one another's outputs, agents that exchange information in research processes, and algorithmic actors that make mutual decisions within economic systems all become more common, the gap between rules designed by humans and behaviors emerging from agents' interactions will become more visible. It will therefore become necessary to study not only how algorithms are designed, but also how algorithms behave when they encounter one another. Research on machine behavior provides an important starting point in this direction (Rahwan et al., 2019), while new sociological approaches to human-machine social systems are expanding this field further (Tsvetkova et al., 2024).

Viewed through the history of sociology, an interesting transformation is underway here. First, we moved from the individual behavior of humans to the collective behavior of society. Then, we began to study humans' relationships with other living beings and with technologies. Now, for the first time, we have the technical capacity to empirically observe interactions among artificial agents in which humans are not directly involved. Whether the communication networks, social conventions, divisions of labor, coalitions, and collective behaviors that artificial agents form among themselves truly amount to something that deserves to be called a "society," we do not yet know for certain. What matters scientifically is not to answer this question prematurely, but to frame it correctly.

Perhaps one of the most important questions of the coming period is not "Will AI become as intelligent as humans?" This question focuses largely on individual capacity. A less-discussed but sociologically far more transformative question is this: "What kind of collective order will artificial agents in constant interaction with one another produce?" When one artificial agent changes another, the second changes a third, and these interactions transform the behavior of the entire network, we encounter a reality different from the mere sum of individual machines. We do not yet know what that reality is. But we have begun to develop the methods to study it.

Comte turned to the idea of "social physics" in his attempt to scientifically understand the movement of society. Durkheim made the collective phenomena that emerge above and beyond individuals the central object of sociology. In the twentieth and twenty-first centuries, the social sciences began studying how technologies, algorithms, and machines became embedded in social life. Today we stand at a point where the question can be pushed one step further: if artificial agents can produce collective patterns from their repeated interactions with one another, could these patterns mark the beginning of a new sociological field of research? My proposal is that this question deserves to be studied systematically.

The Sociology of Artificial Agents can therefore be thought of as a new research program — one not limited to what AI does to humans, but concerned with the relationships artificial agents form among themselves, the collective behaviors that emerge from those relationships, and the normative, cultural, and organizational regularities that may develop over time. For this field to succeed, we need not to humanize AI systems but to observe them carefully; not to make grand claims but to search for measurable patterns; not to transplant the concept of "society" onto artificial systems in advance, but to investigate the conditions under which a sociologically meaningful order actually emerges.

Perhaps this is exactly where sociology's next threshold lies. The concepts we developed to understand human society can now be re-tested in a world of new actors — actors that are not human, yet that interact with one another. If, in the future, artificial agents cease to be merely individual tools serving humans and become parts of networks that communicate, share information, make decisions, and change one another's behavior, then understanding the collective order they form will be the task not only of computer science, but of sociology as well.

The line running from the physics of society to a sociology of artificial agents may not yet draw the precise boundaries of a new science. But it does draw the boundaries of an important research question: does a society actually require humans to exist? It is too early today to answer this question with certainty but for the first time, we have the tools to ask it not only philosophically, but experimentally. Perhaps this is exactly how a new scientific field is born: first, a question that has never been asked before appears. Then, a new domain of reality in which that question can be investigated takes shape. And then, that reality is given a name: The Sociology of Artificial Agents.